# Super-ballistic flow of viscous electron fluid through graphene constrictions

R. Krishna Kumar[1,2,3], D. A. Bandurin[1,2], F. M. D. Pellegrino[4], Y. Cao[2], A. Principi[5], H. Guo[6], G. H. Auton[2], M. Ben Shalom[1,2], L. A. Ponomarenko[3], G. Falkovich[7], I. V. Grigorieva[1], L. S. Levitov[6], M. Polini[1,4,8], A. K. Geim[1,2]

[1]School of Physics & Astronomy, University of Manchester, Manchester M13 9PL, United Kingdom
[2]National Graphene Institute, University of Manchester, Manchester M13 9PL, United Kingdom
[3]Department of Physics, University of Lancaster, Lancaster LA1 4YW, United Kingdom
[4]NEST, Istituto Nanoscienze-CNR and Scuola Normale Superiore, 56126 Pisa, Italy
[5]Radboud University, Institute for Molecules and Materials, 6525 AJ Nijmegen, The Netherlands
[6]Masssachusetts Institute of Technology, Cambridge, Massachusetts 02139, USA
[7]Weizmann Institute of Science, Rehovot 76100, Israel
[8]Istituto Italiano di Tecnologia, Graphene Labs, Via Morego 30, 16163 Genova, Italy

**Electron-electron (e-e) collisions can impact transport in a variety of surprising and sometimes counterintuitive ways[1–6]. Despite strong interest, experiments on the subject proved challenging because of the simultaneous presence of different scattering mechanisms that suppress or obscure consequences of e-e scattering[7–11]. Only recently, sufficiently clean electron systems with transport dominated by e-e collisions have become available, showing behavior characteristic of highly viscous fluids[12–14]. Here we study electron transport through graphene constrictions and show that their conductance below 150 K increases with increasing temperature, in stark contrast to the metallic character of doped graphene[15]. Notably, the measured conductance exceeds the maximum conductance possible for free electrons[16,17]. This anomalous behavior is attributed to collective movement of interacting electrons, which 'shields' individual carriers from momentum loss at sample boundaries[18,19]. The measurements allow us to identify the conductance contribution arising due to electron viscosity and determine its temperature dependence. Besides fundamental interest, our work shows that viscous effects can facilitate high-mobility transport at elevated temperatures, a potentially useful behavior for designing graphene-based devices.**

Graphene hosts a high quality electron system with weak phonon coupling[20,21] such that e-e collisions can become the dominant scattering process at elevated temperatures, $T$. In addition, the electronic structure of graphene inhibits Umklapp processes[15], which ensures that e-e scattering is momentum conserving. These features lead to a fluid-like behavior of charge carriers, with the momentum taking on the role of a collective variable that governs local equilibrium. Previous studies of the electron hydrodynamics in graphene were carried out using the vicinity geometry and Hall bar devices of a uniform width. Anomalous (negative) voltages were observed, indicating a highly viscous flow, more viscous than that of honey[12,22,23]. In this report, we employ a narrow constriction geometry (Fig. 1a) which offers unique insight into the behavior of viscous electron fluids. In particular, the hydrodynamic conductance through such constrictions becomes 'super-ballistic', exceeding the fundamental upper bound allowed in the ballistic limit, which is given by the Sharvin formula[16,17]. This is in agreement with theoretical predictions[18,19] and is attributed to a peculiar behavior of viscous flows that self-organize into streams with different velocities with 'sheaths' of a slow-moving fluid near the constriction edges (Fig. 1b). This cooperative behavior helps charge carriers to circumnavigate the edges and enhances the total conductance. The phenomenon is analogous to the transition from the Knudsen to Poiseuille regimes, well known in gas dynamics, where the hydrodynamic pressure can rapidly drop upon increasing the gas density and the rate of collisions between molecules[24].

Our devices are made of monolayer graphene encapsulated between hexagonal boron nitride crystals as described in Supplementary Section 1. The device design resembles a multi-terminal Hall bar, endowed with constrictions positioned between adjacent voltage probes (Fig. 1c). Below we refer to them as (classical) point contacts (PCs). Five such Hall bars were investigated, each having PCs of various widths $w$ and a reference region without a constriction. The latter allowed standard characterization of graphene, including measurements of its longitudinal resistivity $\rho_{xx}$. All our devices exhibited mobilities exceeding 10 m$^2$ V$^{-1}$ s$^{-1}$ at liquid-helium $T$, which translates into a mean free path exceeding 1 μm with respect to momentum-non-conserving collisions (Supplementary Section 2).

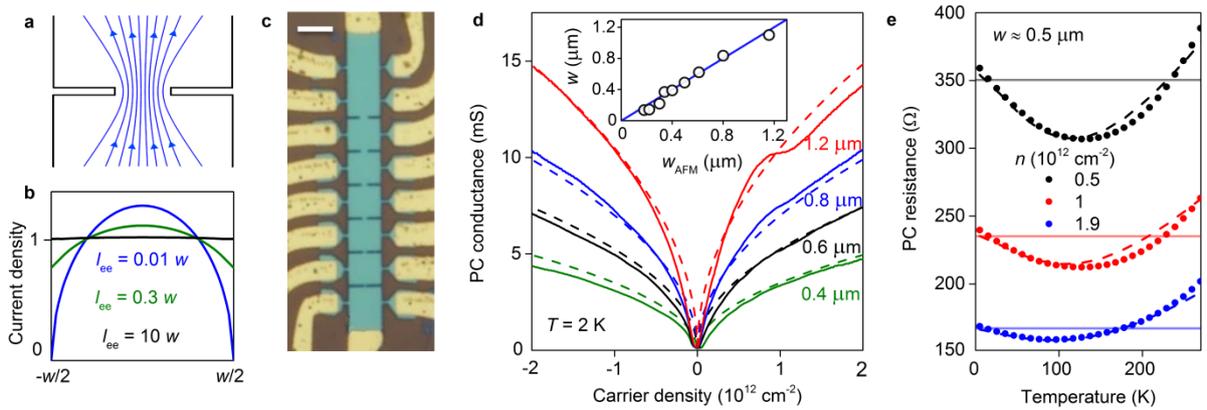

**Figure 1| Electron flow through graphene constrictions. a,** Schematic of viscous flow in a PC. **b,** Distribution of the electric current across the PC, normalized by the total current. In the hydrodynamic regime (e-e scattering length $l_{ee} \ll w$), there is little flow near the edges (blue curve). In the ballistic regime $l_{ee} \gg w$, the current across the aperture is uniform (black curve). **c,** Optical micrograph of one of our devices. Scale bar, 5 μm. The PCs vary in width from 0.1 to 1.2 μm. **d,** Measurements of the low-$T$ conductance for PCs of different $w$ (solid curves). Dashed curves: Ballistic conductance given by eq. (1). Inset: The PC width $w$ found as the best fit to experimental $G_{pc}(n)$ is plotted as a function of $w_{AFM}$. Solid line: $w \equiv w_{AFM}$. **e,** $R_{pc}(T)$ for a 0.5 μm constriction at representative carrier densities. Dots: Experimental data. Horizontal lines: Ballistic resistance given by eq. (1). Dashed curves: Theoretical prediction for our viscous electron fluid, using simplified expressions for $T$ dependence of e-e and electron-phonon scattering ($\propto T^2$ and $T$, respectively). Details are given in Supplementary Section 4.

Examples of the measured PC conductance $G_{pc}$ at 2 K are given in Fig. 1d. In the low-$T$ regime, all scattering lengths exceed $w$ and transport is ballistic, which allows $G_{pc}$ to be described by the Sharvin formula[16]

$$G_b = \frac{4e^2}{h} \frac{w\sqrt{\pi|n|}}{\pi} \qquad (1)$$

where $n$ is the carrier concentration (positive and negative $n$ denote electron and hole doping, respectively). The expression is derived by summing the contributions of individual electron modes that propagate through the constriction with each of them contributing the conductance quantum, $e^2/h$, towards the total conductance. The dashed curves in Fig. 1d show the PC conductance calculated using eq. (1) and assuming the width values, $w_{AFM}$, as determined by atomic force microscopy. The observed agreement between the experiment and eq. (1) does not rely on any fitting parameters. Alternatively, we could fit our experimental curves using eq. (1) and extract the effective width $w$ for each PC (Supplementary Section 3). The results are plotted in the inset of Fig. 1d as a function of $w_{AFM}$. For $w \geq 0.4$ µm, the agreement between $w$ and $w_{AFM}$ is within ~5%. Deviations become larger for our smallest constrictions, suggesting that they are effectively narrower, possibly because of edge defects. Although we focus here on classical PCs with a large number of transmitting modes, we note that our devices with $w < 0.2$ µm exhibit signs of conductance quantization, similar to those reported previously[25,26].

The central result of our study is presented in Fig. 1e. It shows that the resistance of graphene PCs, $R_{pc} \equiv 1/G_{pc}$, is a *non-monotonic* function of $T$, first decreasing as temperature increases. This behavior, typical for insulators, is unexpected for our metallic system. It is also in contrast to the $T$ dependence of $\rho_{xx}$ observed in our Hall bar devices. They exhibit $\rho_{xx}$ monotonically increasing with $T$, the standard behavior in doped graphene (Supplementary Section 2). All our PCs with $w < 1$ µm exhibited this anomalous, insulating-like $T$ dependence up to 100–150 K (Fig. 2a). As a consequence, $G_{pc}$ in its maximum could exceed the ballistic limit value by > 15% (Fig. 1e). At higher $T$, $R_{pc}$ starts growing monotonically and follows the same trend as $\rho_{xx}$. The minima in $R_{pc}(T)$ were more pronounced for narrower constrictions (Fig. 2a), corroborating the importance of the geometry. Figures 2b-c elaborate on the non-metallic behavior of graphene PCs by plotting maps of the derivative $dR_{pc}/dT$ as a function of both $n$ and $T$. The anomalous insulating-like $T$ dependence shows up as the blue regions whereas the metallic behavior appears in red. For narrow constrictions, the anomalous behavior was observed for all accessible $n$ below 100 K, becoming most pronounced at low densities but away from the neutrality point (Figs. 1e, 2b). For wide PCs (Fig. 2c), the non-metallic region becomes tiny, in agreement with the expected crossover from the PC to standard Hall bar geometry.

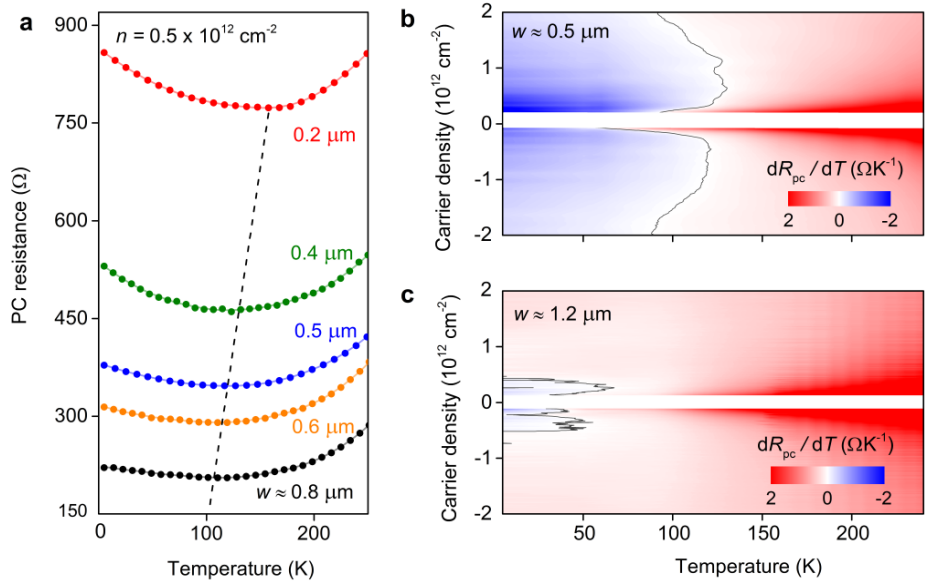

**Figure 2| Transition from metallic to insulating behavior in constrictions of different widths. a,** Temperature dependence for PCs with different $w$ at a given $n$. The dashed line indicates that the minima shift to higher $T$ and become deeper for narrower constrictions. **b-c,** Color map $dR_{pc}/dT(T,n)$ for $w \approx 0.5$ and 1.2 µm. The black contours mark a transition from the negative to positive $T$ dependence. The white stripes near zero $n$ cover regions near the neutrality point, in which charge disorder becomes important and transport involves thermal broadening and other interaction effects[12,13] beyond the scope of this work.

To describe the non-metallic behavior in our PCs, we first invoke the recent theory[18] that predicts that e-e scattering modifies eq. (1) as

$$G = G_b + G_\nu \quad \text{where} \quad G_\nu = \frac{\sqrt{\pi|n|}\, e^2\, w^2\, v_F}{32\, \hbar\, \nu}, \quad (2)$$

$v_F$ is the Fermi velocity and e-e collisions are parameterized through the kinematic viscosity $\nu = v_F l_{ee}/4$. The quantity $G_\nu$ is calculated for the Stokes flow through a PC in the extreme hydrodynamic regime (that is, for the e-e scattering length $l_{ee} \ll w$). The additive form of Eq. (2) is valid[18,19] for all values of $l_{ee}/w$, even close to the ballistic regime $l_{ee} \gg w$. This implies that $G$ should increase with $T$ (in the first approximation[15,27], as $\propto 1/l_{ee} \propto T^2$), which leads to the insulating-like behavior. Eq. (2) also suggests that the viscous effects should be more pronounced at low $n$ where electron viscosity is smaller, in agreement with the experiment (Figs. 1e, 2b). The description by eq. (2) is valid until phonon scattering kicks in at higher $T$. To describe both low-$T$ and high-$T$ regimes on an equal footing, we extended the transport model of ref. 18 to account for acoustic-phonon scattering using an additional term $\propto T$ in the kinetic equation (Supplementary Section 4). The results are plotted in Fig. 1e showing good qualitative agreement with the experiment.

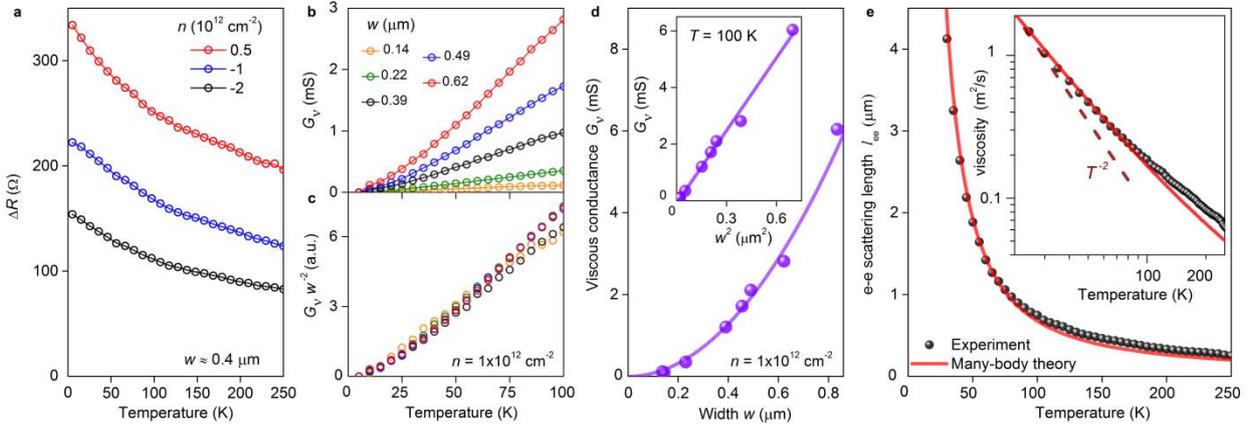

**Figure 3| Quantifying e-e interactions in graphene. a,** $T$ dependence of the PC resistance after subtracting the contribution from contact regions. **b,** Viscous conductance $G_\nu$ at a given $n$ for PCs with $w$ ranging between 0.1 and 0.6 µm. **c,** Data from (b) normalized by $w^2$. **d,** $G_\nu$ as a function of $w$ for given $T = 100$ K and $n = 10^{12}$ cm$^{-2}$. Solid curve: Best fit to eq. (2) yields $\nu \approx 0.16$ m$^2$ s$^{-1}$, a value 5 orders of magnitude larger than the viscosity of water. Inset: Same data as a function of $w^2$. **e,** $T$ dependence of the e-e scattering length found as $l_{ee} = 4\nu/v_F$ (symbols) for $n = 10^{12}$ cm$^{-2}$ and $w \approx 0.5$ µm. Red curve: Microscopic calculations of $l_{ee}$ (Supplementary Section 6). Inset: $\nu(T)$ on a log-log scale. The data are from the main panel and color-coded accordingly. The dashed line indicates the $1/T^2$ dependence.

For further analysis, we used our experimental data to extract $G_\nu$, which in turn enabled us to determine $\nu$ and $l_{ee}$. To this end, we first followed the standard approach in analysis of transport data for quantum PCs, which takes into account the contact resistance $R_C$ arising from the wide regions leading to constrictions[17,28]. Accordingly, the total resistance of PCs can be represented as

$$R_{pc} = (G_b + G_\nu)^{-1} + R_C \quad (3).$$

To avoid fitting parameters, we model the contact resistance as $R_C = b\rho_{xx}$ where $b$ is a numerical coefficient calculated by solving the Poisson equation for each specific PC geometry and $\rho_{xx}$ is taken as measured from the reference regions. For our devices, $b$ ranged between 2 and 5 (Supplementary Section 5). Examples of the resulting $\Delta R = R_{pc} - b\rho_{xx}$ are plotted in Fig. 3a. The figure shows that, after the rising phonon contribution is accounted for through $R_C$, the resistance attributable to the narrowing itself monotonically decreases with increasing $T$ over the entire $T$ range, in agreement with eq. (2). As a next step, we use the conductance found in the limit of low $T$ as $G_b$ for each PC and subtract this value to find the viscous conductance $G_v$. The results are shown in Fig. 3b for several PCs. Remarkably, if $G_v$ is normalized by $w^2$, all the experimental data collapse onto a single curve (Fig. 3c). This scaling is starkly different from the Sharvin dependence $G_b \propto w$ observed in the ballistic regime (Fig. 1d) and, more generally, from any known behavior of electrical conductance that always varies linearly with the sample width. However, our result is in excellent agreement with eq. (2) that suggests $G_v \propto w^2$. The $w^2$ scaling behavior is further validated in Fig. 3d, lending strong support to our analysis.

The measured dependence $G_v(T)$ allows us to extract $\nu(T)$ and $l_{ee}(T)$ for graphene using eq. (2). The results are shown in Fig. 3e and compared with the calculations[29] detailed in Supplementary Section 6. The agreement is surprisingly good (especially taking into account that neither experiment nor calculations use any fitting parameters) and holds for different PC devices and different carrier densities (Supplementary Section 7). We also note that the agreement is considerably better than the one achieved previously using measurements of $\nu$ in the vicinity geometry[12] and even accommodates the fact that both experimental and theoretical curves in Fig. 3e (inset) deviate from the $1/T^2$ dependence expected for the normal Fermi liquid[6,30]. The deviations arise because temperatures ~50–100 K are not insignificant with respect to the Fermi energy. Furthermore, our calculations in Fig. 3e stray slightly off the experimental curve above 100 K. In fact, this is expected because, in the hydrodynamic regime $l_{ee} \ll w$, the kinematic viscosity can no longer be expressed in terms of $l_{ee}$ (as above) and requires a more accurate description using the two-body stress-stress response function[29]. Although the strong inequality $l_{ee} \ll w$ is not reached in our experiments, the experimental data in Fig. 3e do tend in the expected direction (Supplementary Section 8).

To conclude, graphene constrictions provide a unique insight into the impact of e-e interactions on electron transport. The observed negative $T$ dependence of the point contact resistance, its super-ballistic values and the unusual $w^2$ scaling are clear indicators of the important role of e-e collisions in graphene at elevated temperatures. Our analysis, based on the experimental measurements and microscopic calculations, offers a guide for disentangling intriguing phenomena at the crossover between the ballistic and hydrodynamic transport regimes.

## Supplementary Information

### S1. Device fabrication

Our encapsulated-graphene devices were made following a recipe similar to that used in the previous reports[1,2,3]. First, an hBN-graphene-hBN stack was assembled using the dry peel technique[2]. This involved mechanical cleavage to obtain monolayer graphene and hBN crystals less than 50 nm thick. The selected crystallites were stacked on top of each other using a polymer membrane attached to a micromanipulator[2]. The resulting heterostructure was transferred on top of an oxidized silicon wafer (290 nm of $SiO_2$) which served in our experiments as a back gate. After this, the heterostructure was patterned by electron beam lithography to first define contact regions. Reactive ion etching (RIE) was employed to selectively remove the heterostructure areas unprotected by the lithographic mask, which resulted in trenches for depositing long electrical leads and metal contacts to graphene (Fig. S1a). 3 nm of chromium followed by 80 nm of gold were evaporated into the trenches. This fabrication sequence allowed us to prevent contamination of the narrow graphene edges that were exposed by RIE, which reduced the contact resistance[3].

Next, the same lithography and etching procedures were employed again to define the final device geometry. Figure S1a shows another device used in our experiments (in addition to that shown in Fig. 1c of the main text). The two Hall bars host four constrictions and an accompanying reference region. To determine their width, point contacts (PCs) were imaged by atomic force microscopy (AFM). An example of the obtained AFM images is provided in Fig. S1b, and a line trace in Fig. S1b shows a typical height profile $h(x)$ across the constriction. Because of much quicker etching of hBN in comparison with graphene, a step-like feature develops in the etched slope[3] as indicated by the arrow in Fig. S1b. This feature allows us to accurately determine the vertical position of the graphene channel. To calculate its width $w_{AFM}$, we took into account both graphene's vertical position (Fig. S1c) and a finite opening angle of our AFM tips (~20°).

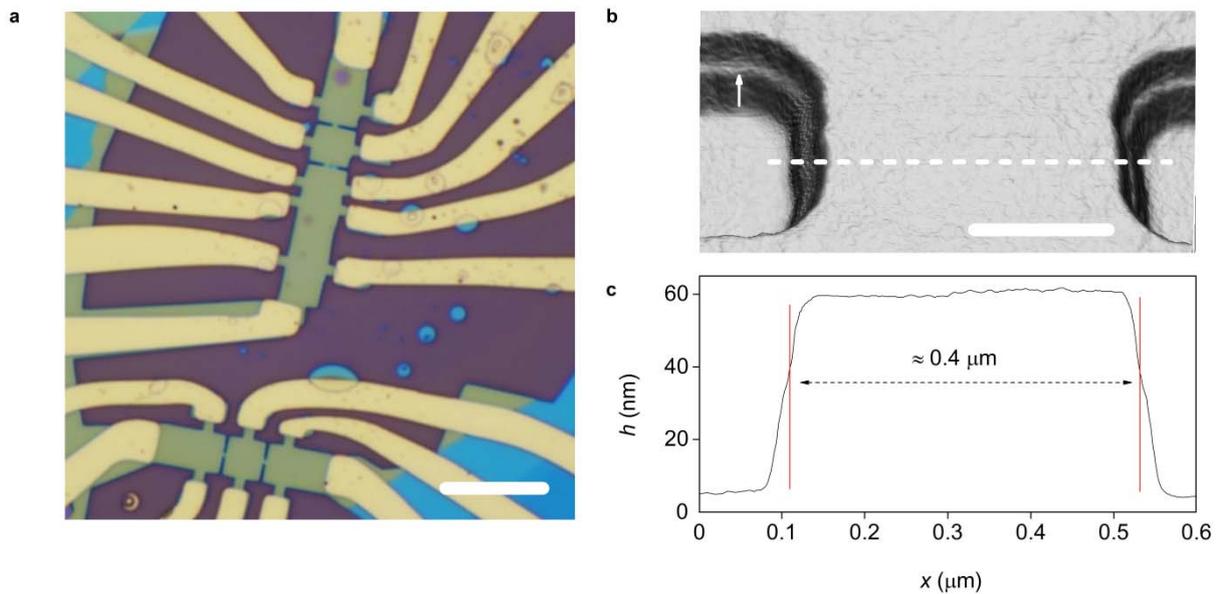

**Figure S1 | Graphene point contacts. a,** Optical image of a device with PCs varying in width from 0.2 to 0.6 μm. Scale bar: 10 μm. **b,** Three dimensional AFM image of one of the point contacts. Scale bar: 0.2 μm. **c,** Height profile along the white dashed line in (b). Red lines indicate the width $w_{AFM}$ for this particular constriction; graphene is buried 20 nm under the hBN layer.

## S2. Mobility and mean free path

We characterized quality of our graphene devices using their reference regions. The longitudinal and Hall resistivities ($\rho_{xx}$ and $\rho_{xy}$, respectively) were measured in the standard four-probe geometry as a function of back gate voltage. Figure S2a shows $\rho_{xx}(n)$ at different $T$, where carrier density $n$ was determined from $\rho_{xy}$. One can see a typical behavior for high quality graphene. At low $T$, $\rho_{xx}$ exhibits a peak at the charge neutrality point (NP) with a sharp decrease down to 20-50 $\Omega$ for $|n| > 0.5 \times 10^{12}$ cm$^{-2}$. Away from the NP, $\rho_{xx}$ grows monotonically with $T$ (inset of Fig. S2a) as expected for phonon-limited transport in doped graphene[4].

The mobility was calculated using the Drude formula, $\mu = 1/ne\rho_{xx}$ where $e$ is the electron charge. For typical $n \sim 1 \times 10^{12}$ cm$^{-2}$, $\mu$ exceeded 15 m$^2$V$^{-1}$s$^{-1}$ at 5 K and was around 5 m$^2$V$^{-1}$s$^{-1}$ at room temperature. These values translate into the elastic mean free path $l = \mu\hbar/e(n\pi)^{0.5}$ of about 1 to a few microns at all $T$ (Fig. S2b) which exceeds the dimensions of our graphene PCs and implies ballistic transport through the constrictions with respect to momentum-non-conserving collisions. To illustrate that such ballistic transport occurs not only inside reference regions but also for the sections of our devices with PCs, we carried out measurements in the bend geometry[5,6] (micrograph in Fig. S2c). This figure shows an example of the bend resistance $R_B(n)$ measured from a region located between two PCs. For $n$ away from the NP, $R_B$ becomes negative, which indicates direct, ballistic transmission of charge carriers from, for example, current contact (1) into voltage contact (4) (refs. 5,6). The negative bend resistance was found for all the regions of our devices, proving their high homogeneity and, also, implying that $l$ at low $T$ was at least 4 μm (our Hall bars' width), somewhat higher than the above estimates based on the Drude model (inset of Fig. S2b).

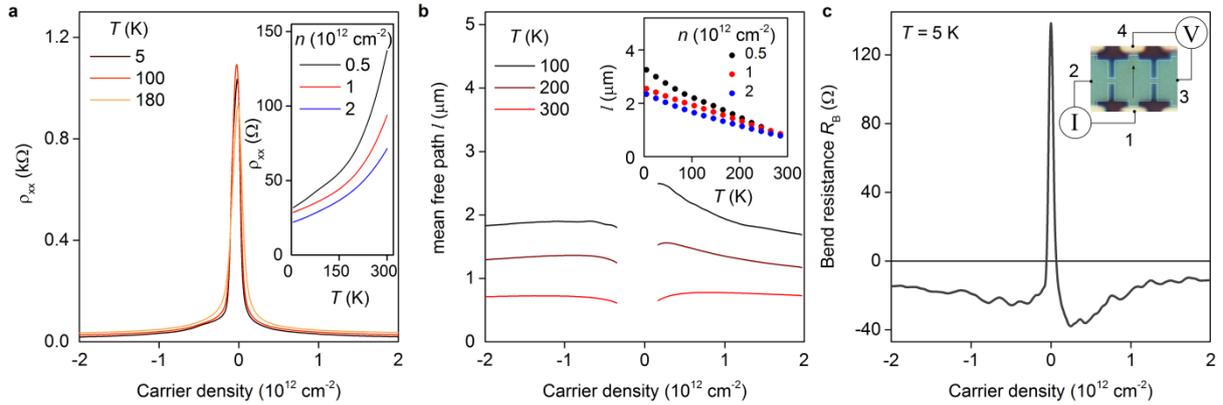

**Figure S2|** **Characterization of encapsulated graphene. a**, $\rho_{xx}$ as a function of $n$ at different temperatures. Inset: $\rho_{xx}(T)$ for a few $n$. **b**, Elastic mean free path as a function of $n$ at high $T \geq 100$ K. Inset: Complete $T$ dependence for various $n$. **d**, Bend resistance $R_B(n)$ at low $T$. The micrograph shows schematics the bend geometry used in the experiment where $R_B = R_{12,34}$ (for details see refs. 5,6).

## S3. Finding the width of point contacts

In a conventional two-dimensional electron gas (e.g., in GaAlAs heterostructures), local gates are used to deplete charge carriers in specific areas, creating insulating regions that inhibit current pathways. This allows constrictions with smooth edges. In graphene devices, constrictions are made by milling away the material. Accordingly, our PCs are defined by actual graphene edges. Figures S3a-b show two more examples of AFM images of our PCs with $w_{AFM} \approx 0.2$ and 0.5 μm. Due to limitations of electron-beam lithography, the edge profiles are unavoidably rough on a sub-100-nm scale. The destructive nature of RIE may also introduce microscopic cracks[7] that cannot be visualized being buried under the top hBN layer. Such edge disorder may be responsible for the lowering of the PC conductance below the Sharvin limit[7] and is expected to contribute more in our narrowest devices (Fig. 1d of the main text).

To gain further information about our narrowest PCs, we compared their measured conductance with that expected from the Sharvin formula. Figure S3d mirrors the presentation in Fig. 1d of the main text, showing the PC conductance as a function of density $n$, for the constrictions presented in Figs. S3a-c. The theory curves are again plotted using the width measured by AFM. In the case of $w_{AFM} \approx 0.2$ μm, $G_{pc}$ was found to be notably lower than that expected from eq. (1) of the main text. As discussed above, this can be attributed to the edge roughness playing a relatively more prominent role for narrower constrictions[7]. However, even for the narrowest PC, its $G_{pc}(n)$ still scales linearly with the Fermi wave vector $k_F$, following the Sharvin formula (inset of Fig. S3d). This allows us to find the constriction's effective width $w$. We used such linear fits to determine effective widths for all our PC devices. Figure S3e shows examples of the fitting procedure for five PCs, plotting $G_{pc}$ as a function of $k_F$. In all our devices, the dependences $G_{pc}(k_F)$ were clearly linear which shows that the effective width $w$ is a good approximation for describing graphene constrictions. Such an approach was also used previously for suspended graphene constrictions[8].

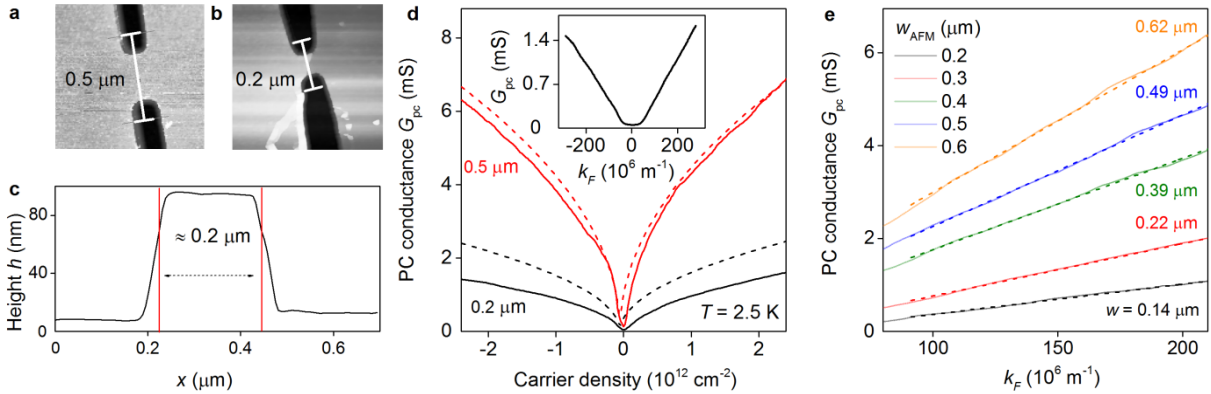

**Figure S3| Point contact widths. a-b**, AFM images of our constrictions. Grey scale: black - 0 nm; white - 95 nm. **c**, Height profile across the narrowest constriction, similar to the presentation in Fig. S1. **d**, Low-$T$ conductance for the devices in (a) and (b). Solid curves: Experimental data. Dashed: Sharvin expression using the width determined by AFM. Inset: $G_{pc}$ for the 0.2 μm PC is re-plotted as a function of $k_F$. **e**, $G_{pc}$ as a function of $k_F$ for several PCs measured at 2 K (electron doping). The dashed lines are linear fits to our experimental data (solid curves). The effective width $w$, extracted from the best fits to eq. (1) of the main text, is color-coded for each constriction.

**S4. Modelling the ballistic-to-viscous crossover**

Transport measurements reported in the main text were carried out using constrictions with $w$ ranging from 0.2 to 1.2 μm and carrier densities of the order of $10^{12}$ cm$^{-2}$. The observed 'super-ballistic' behavior (that is, the suppression of the PC resistance below the ballistic Sharvin-Landauer value) was found to be most prominent at temperatures below 100 K. Under these conditions the e-e scattering mean free path $l_{ee}$, which depends on $T$ and $n$, is comparable to the constriction width $w$. Therefore, modelling electron transport in our experimental system requires a method that can

S3

operate at the crossover between the ballistic and hydrodynamic regimes. To this end, we have used an approach developed in ref. 9, which is based on a kinetic equation with the collision operator describing momentum-conserving e-e collisions. In the absence of momentum-relaxing processes, such as electron-phonon scattering, this approach predicts the conductance $G_{pc}$ that attains a ballistic value at zero $T$ and increases monotonically with increasing temperature. In the present work, to account for the non-monotonic temperature dependence of the measured resistance $R_{pc}$, first growing and then decreasing, we have extended the model of ref. 9 by adding to the kinetic equation a momentum-relaxing term that describes electron-phonon scattering. In notations of ref. 9 our model reads

$$(\partial_t + v\nabla_x)f(\theta,x) = -\gamma_{ee}(1-P)f(\theta,x) - \gamma_{ep}(1-P_0)f(\theta,x). \tag{S1}$$

where $f(\vartheta,\mathbf{x})$ is the non-equilibrium carrier distribution at the 2D Fermi surface parameterized by the angle $\vartheta$. The rates $\gamma_{ee}$ and $\gamma_{ep}$ describe the e-e scattering and electron-phonon scattering processes, the quantities $P$ and $P_0$ are projectors on the angular harmonics with $m = 0, \pm1$ and $m = 0$, respectively, and 1 stands for the identity operator. As in ref. 9, this model assumes that all harmonics of the distribution function, which are not conserved, should relax at equal rates. The relaxation rates are equal to $\gamma_{ee} + \gamma_{ep}$ for $m = \pm2, \pm3,...$ and $\gamma_{ee}$ for $m = \pm1$. The single-rate assumption allows us to reduce the integral-differential kinetic equation to a closed-form self-consistency relation for quasi-hydrodynamic variables (i.e., the $m = 0, \pm1$ angular harmonics), providing a means for solving it in the constriction geometry.

Incorporating the electron-phonon scattering term in the approach of ref. 9 significantly changes the algebra but conceptually proves to be uneventful. Given the scattering rate values $\gamma_{ee}$ and $\gamma_{ep}$, we first find the current profile in the constriction cross-section. This is done considering non-slip boundary conditions, which we modelled by adding to the right-hand side of eq. (S1) a delta function term of the form $-b\delta(y)\delta(w/2 - |x|)P_\pm f(\theta,x)$ where the operator $P_\pm$ projects $f(\theta,x)$ on the $m = 0,\pm1$ angular harmonics. The parameter $b$ is taken to the limit $b \to \infty$ to model an impenetrable boundary at the half-lines $y = 0$, $|x|>w/2$. We then derive a self-consistent relation for current density in the constriction, solve it numerically and use the solution to determine the potential distribution in the regions adjacent to the constriction. The potential difference, obtained for the unit total current, yields the resistance.

As a simple model, we use the temperature dependences for the rates $\gamma_{ee}$ and $\gamma_{ep}$ in the following form

$$\gamma_{ee} = \frac{aT^2}{n^{1/2}} \times \frac{v_F}{w}, \qquad \gamma_{ep} = cT \times \frac{v_F}{w} \tag{S2}$$

where $v_F = 10^6$ m/s is the graphene Fermi velocity. These dependences correspond to the prediction of the Fermi liquid theory at weak coupling and the electron-phonon scattering rate due to acoustic phonons. The fits to the experimental dependences $R_{pc}(T)$ shown in Fig. 1e of the main text were obtained with the best-fit values of $a = 8.6\times10^3$ K$^{-2}$m$^{-1}$ and $c = 2\times10^{-3}$ K$^{-1}$, which were taken to be identical for all densities $n$. To test the robustness of our model, we also explored other power-law and polynomial temperature dependences, and found that modest deviations from the $T^2$ and $T$ scaling do not impact quality of the fits and may even lead to slight improvement. The agreement between the fits and the experimental data in Fig. 1e, impressive as it is, should therefore not be taken as evidence for the $T^2$ and $T$ scaling for the rates $\gamma_{ee}$ and $\gamma_{ep}$. Indeed, the analysis presented in the final part of the main text effectively uses a faster $T$ dependence for phonon scattering and $\gamma_{ee}$ somewhat slower than $T^2$, which provides a surprising good quantitative agreement with the experimental data.

## S5. Ohmic contribution to point contact resistance

Narrow constrictions that define PCs are connected to broader regions in which current and voltage contacts are located (see the above images of our experimental devices). In the presence of elastic scattering, these regions are responsible for an additional Ohmic contribution $R_C$ that depends on details of device's geometry. The contribution grows with increasing temperature (that is, with increasing electron-phonon scattering) and can obscure the viscous-flow behavior, as discussed in the main text.

To account for the contribution from the contact regions, we calculated $R_C$ in the limit of diffusive transport and then subtracted the obtained value from the measured resistance $R_{pc}$. To this end, we computed $R_C = V_{12}/I_{56}$ (see Fig. S4a) by solving the following set of equations

$$\nabla \cdot J(r) = 0, \quad \frac{\sigma_0}{e} \nabla \phi(r) - J(r) = 0 \tag{S3}$$

where $J(r)$ is the current density, $\phi(r)$ is the electric potential in the two-dimensional plane and $\sigma_0 = ne^2\tau/m$ is the Drude-like conductivity with $m$ and $e$ being the effective mass and the electron charge, respectively. To solve the above differential equations, we followed the procedure used in ref. 10. In brief, by discretizing the differential operators on a square mesh, we obtained a set of sparse linear equations that could readily be solved. Our method involved three different staggered meshes that sampled values of the potential and, independently, the two components of the current density[10]. This was required to ensure that the velocity component orthogonal to the boundary was sampled, too. Finally, we used the following boundary conditions to simulate device's edges and contacts: (i) the current orthogonal to the edges was zero, (ii) the current was also zero through voltage contacts, (iii) the total current through source and drain contacts was fixed, as in the experiment.

Exploiting the linearity of the problem, we can write the Ohmic contribution as $R_C = b\rho_{xx}$, where $b$ is a dimensionless function of the ratios $w/W$ and $L/W$. The calculated coefficient $b$ is plotted in Fig. S4b as a function of $w/W$ for the geometry used in our experiments with $L/W = 1$.

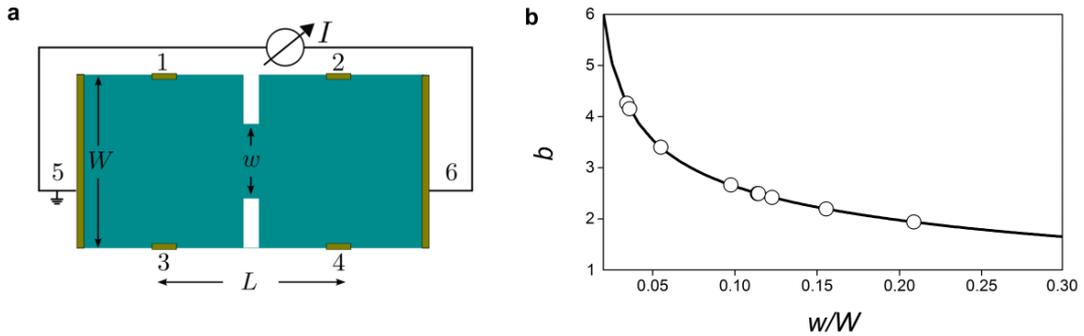

**Figure S4| Ohmic contribution**. **a**, Schematic of the device geometry. Electrical current $I$ is passed between contacts 5 and 6. Voltage drop is measured between pairs of contacts 1 and 2 or 3 and 4. **b**, Coefficient $b$ as a function of $w/W$ for the given $L/W = 1$. The solid curve shows our numerical results. The open circles correspond to the geometry of PC devices measured in this work.

## S6. Microscopic calculations of electron-electron scattering

In this section, we provide details of microscopic calculations of $l_{ee}$ which were presented in Fig. 3e of the main text. We have determined $l_{ee}$ from the imaginary part of the retarded quasiparticle self-energy $\Sigma_\lambda(k, \omega)$ averaged over the Fermi surface[11]. The conduction and valence bands are marked with $\lambda = +$ and $-$, respectively. For an electron-doped system, we use

S5

$$l_{ee}^{-1} \equiv \frac{2}{v_F} \int d\omega \left(\frac{\partial n_F(\omega)}{\partial \omega}\right) \Im m[\Sigma_+(k_F, \omega)] \tag{S4}$$

where $k_F$ is the Fermi wave vector and $n_F(\omega)$ is the Fermi distribution. Below we use $\hbar \equiv 1$ and $k_B \equiv 1$ for the Planck and Boltzmann constants, respectively. In the spirit of the large-$N$ approximation (where $N = 4$ is the number of fermion flavors in graphene), the quasiparticle self-energy $\Sigma_\lambda(k, \omega)$ can be calculated within the $G_0W$ approximation. For monolayer graphene[12,13]

$$\Im m[\Sigma_\lambda(k_F, \omega)] = \int \frac{d^2\boldsymbol{q}}{(2\pi)^2} \sum_{\lambda'} \Im m[W(q, \omega - \xi_{\boldsymbol{k}-\boldsymbol{q},\lambda'})] \mathcal{F}_{\lambda\lambda'}(\theta_{\boldsymbol{k},\boldsymbol{k}-\boldsymbol{q}}) [n_B(\omega - \xi_{\boldsymbol{k}-\boldsymbol{q},\lambda'}) + n_F(-\xi_{\boldsymbol{k}-\boldsymbol{q},\lambda'})] \tag{S5}$$

where $n_{F/B}(\varepsilon) = (e^{\varepsilon/T} \pm 1)^{-1}$ are the usual Fermi and Bose distribution factors, respectively, and $W(q,\omega) = V(q,\omega)/\varepsilon(q,\omega)$ is the screened Coulomb interaction. The Fourier transform of the bare Coulomb interaction, $V(q,\omega) = 2\pi e^2 \mathcal{G}(qd,qd')/q$, contains the form-factor $\mathcal{G}(qd,qd')$, which encodes all the information about the dielectric environment surrounding the graphene. It depends on the thickness $d$ and $d'$ of hBN above and below the graphene plane, as well as on the in-plane $\epsilon_x$ and out-of-plane $\epsilon_z$ components of the dielectric tensor of hBN. The full expression for $\mathcal{G}$ is given, for example, in the Supplementary Material of ref. 14. Finally, $\xi_{\boldsymbol{k},\lambda} = \lambda v_F k - \mu(T)$ is the band energy measured from the chemical potential $\mu(T)$ and $\varepsilon(q,\omega) = 1 - V(q,\omega)\chi_{nn}(q,\omega)$ is the RPA dynamical dielectric function. Here, $\chi_{nn}(q,\omega)$ is the density-density response function of graphene, which can be found in refs. 15–19. $\mathcal{F}_{\lambda\lambda'}(\theta_{\boldsymbol{k},\boldsymbol{k}-\boldsymbol{q}}) = [1 + \lambda\lambda'\cos(\theta_{\boldsymbol{k},\boldsymbol{k}-\boldsymbol{q}})]/2$ is the square of the matrix element of the density operator, with $\theta_{\boldsymbol{k},\boldsymbol{k}-\boldsymbol{q}} = \theta_{\boldsymbol{k}} - \theta_{\boldsymbol{k}-\boldsymbol{q}}$ being the angle between the vectors $\boldsymbol{k}$ and $\boldsymbol{k} - \boldsymbol{q}$.

For completeness, we note that in the Fermi liquid regime[12] eq. (S4) can be simplified to

$$l_{ee}^{-1} = \frac{\pi k_F}{N} \left(\frac{T}{\varepsilon_F}\right)^2 \ln\left(\frac{2\varepsilon_F}{T}\right) \tag{S6}$$

where $\varepsilon_F = v_F k_F$ is the Fermi energy.

### S7. Sample and density dependences of e-e scattering length

In monolayer graphene, where charge carriers are massless Dirac fermions, e-e scattering is dominated by processes that transfer a small amount of the momentum[13]. Such events, usually referred to as collinear collisions, are weakly sensitive to the dielectric enviroment[17]. Therefore, our devices with different thicknesses of top and bottom hBN layers are not expected[13] to exhibit drastically different $l_{ee}$. Indeed, Fig. S5a plots $l_{ee}(T)$ for several PCs in two of our devices with different $d$ and $d'$. For these devices, the e-e scattering lengths calculated as described in Section 6 are indistinguishable on the scale of Fig. S5a, yielding the same curve. As for the experiment, $l_{ee}$ found for all our PCs closely follow the same functional dependence (see Fig. S5a) and exhibit quantitative agreement with the calculations. This substantiates the robustness of the experimental and analytical methods used in this report.

Until now, we presented $l_{ee}(T)$ only for fixed carrier densities. For completeness, Fig. S5b shows the density dependence of $l_{ee}$ at fixed $T$. To find $l_{ee}(n)$, we followed the same analytical procedure as explained in the main text, which allowed us to extract the viscous conductance $G_v$ and, consequently, obtain $l_{ee}$ without using any fitting parameters. Comparison in Fig. S5b between our experiment and calculations again shows good agreement. Perhaps unsurprisingly, it holds best for intermediate $T$ around 100 K, where our PCs are sufficiently away from the purely ballistic regime while the electron-phonon contribution to $R_{pc}$ remains relatively small. Let us note that, in this experiment, $l_{ee}$ slowly increases with $n$, which is in contrast to the trend reported for the vicinity geometry (see Ref. 12 of the main text) but in agreement with the theory that expects $l_{ee}$ to be approximately proportional to $n^{0.5}$.

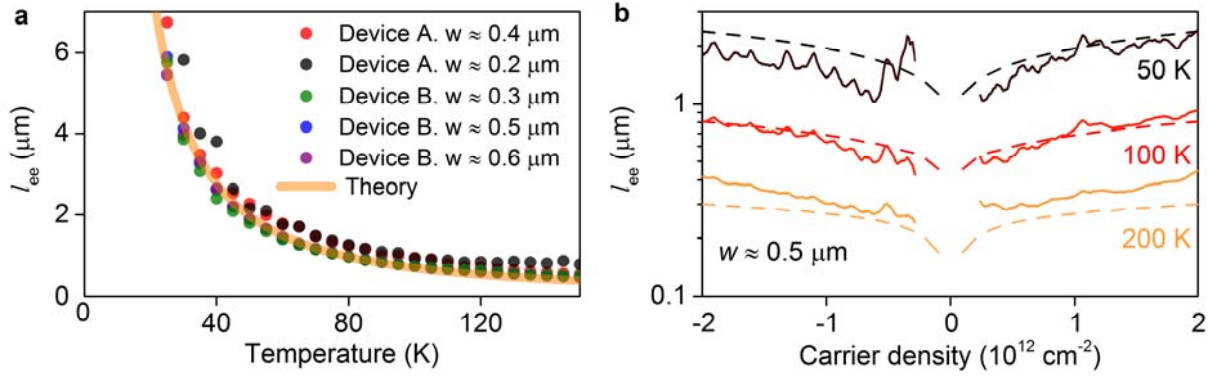

**Figure S5 | Electron-electron scattering for different devices and carrier densities. a,** $l_{ee}$ as a function of $T$ measured using devices A and B with several PCs; $n = 1\times10^{12}$ cm$^{-2}$. Device A is made of graphene encapsulated between hBN crystals off approximately equal thickness ($d \approx d' \sim 40$ nm). In device B, top hBN is ~20 nm whereas the bottom one ~30 nm. Orange curve: Microscopic calculations of $l_{ee}(T)$ for both A and B. **b,** $l_{ee}$ as a function of $n$ at different $T$ in a constriction with $w \approx 0.5$ μm (solid curves). Dashed curves: Calculations of $l_{ee}(n)$.

### S8. Different length scales for electron viscosity

Our experimental data allow us to determine the characteristic length for e-e collisions responsible for the super-ballistic flow. As discussed above and in the main text, we find that these lengths agree extremely well with the e-e mean free path $l_{ee}$, associated with the quasiparticle lifetime $\tau_{ee} = l_{ee}/v_F$. However, at high temperatures, deep in the hydrodynamic regime, the quasiparticle lifetime is expected to be no longer the relevant length scale governing the viscous electron flow. In this regime, the kinematic viscosity ν is better described by the 'viscous' mean free path $l_V$, which is of the same order but not identical to $l_{ee}$.

The kinematic viscosity ν is related to $l_V$ by the standard expression $\nu = v_F l_V/4$ and can be calculated from the stress-stress linear response function $\chi_{ij,kl}(\mathbf{q}, \omega)$ as

$$\nu = -\lim_{\omega\to 0}\frac{1}{4\,n\,m_c\omega}\sum_{i,j=x,y}\Im m\left[\chi_{ij,ij}(\mathbf{0},\omega) - \frac{1}{2}\chi_{ii,jj}(\mathbf{0},\omega)\right], \tag{S7}$$

where $m_c = k_F/v_F$ is the effective mass for monolayer graphene. After rather lengthy calculations (see ref. 20 for technical details), the viscosity length is found to be given by

$$\ell_v^{-1} = \frac{2}{v_F}\int d\omega \left(\frac{\partial n_F(\omega)}{\partial \omega}\right)\Im m\left[\Sigma_+^{(v)}(k_F,\omega)\right], \tag{S8}$$

where

$$\Im m\left[\Sigma_\lambda^{(v)}(k_F,\omega)\right]$$
$$= \int \frac{d^2\mathbf{q}}{(2\pi)^2}\sum_{\lambda'}\Im m[W(q,\omega-\xi_{\mathbf{k-q},\lambda'})]\mathcal{F}_{\lambda\lambda'}(\theta_{\mathbf{k,k-q}})[n_B(\omega-\xi_{\mathbf{k-q},\lambda'})$$
$$+ n_F(-\xi_{\mathbf{k-q},\lambda'})]\sin^2(\theta_{\mathbf{k,k-q}}) \tag{S9}.$$

In the Fermi liquid regime[12] the viscosity length behaves as

$$\ell_v^{-1} = \mathcal{N}\alpha_{ee}^2 k_F \left(\frac{T}{\varepsilon_F}\right)^2, \tag{S10}$$

where $\alpha_{ee}$ = 2.2 is the e-e coupling constant of graphene, and the coefficient $\mathcal{N} \sim 0.1$ has a rather cumbersome expression, depending on microscopic details (see ref. 20).

Figure S6 compares our experimental data (same as in Fig. 3e in the main text) with microscopic calculations for both lengths $l_{ee}$ and $l_v$. As shown in the main text, the experimental data follows $l_{ee}$ closely until about 100 K. Beyond this $T$, the extracted length deviates slightly upwards from $l_{ee}$ and tends towards $l_v$ as expected in the extreme hydrodynamic regime $l_{ee} \ll w$. Proper validation of this transition from $l_{ee}$ to $l_v$ would require measurements at much higher $T$, inaccessible for our experimental devices. Accordingly, Fig. S6 is used here only to point out similarities and differences between our experimental data and e-e scattering length scales, whilst better theoretical understanding is required to make any further conclusions.

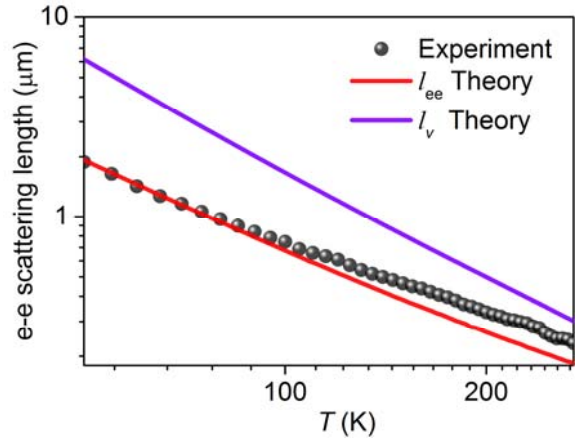

**Figure S6| Different viscous length**. Black symbols: Electron-electron scattering length determined experimentally for a graphene constriction with $w \approx 0.5$ um; $n = 10^{12}$ cm$^{-2}$. Red and purple curves: Microscopic calculations of $l_{ee}$ and $l_v$ as a function of $T$ for the given $n$.

**Supplementary References**